%% file: main.tex
\documentclass[conference]{IEEEtran}
\IEEEoverridecommandlockouts
\usepackage{cite}
\usepackage{amsmath,amssymb,amsfonts}
\usepackage{algorithmic}
\usepackage{graphicx}
\usepackage{textcomp}
\usepackage{xcolor}
\usepackage{subcaption}
\def\BibTeX{{\rm B\kern-.05em{\sc i\kern-.025em b}\kern-.08em
    T\kern-.1667em\lower.7ex\hbox{E}\kern-.125emX}}
\begin{document}

\title{Multilingual Agent System\\for Inclusive Wildfire Evacuation Guidance}

\input{authors}

\maketitle

\input{abstract}


\input{introduction}
\input{system_overview}
\input{results}
\input{conclusion}
\bibliographystyle{IEEEtran}
\bibliography{references}

\end{document}

%% file: authors.tex
\author{
\IEEEauthorblockN{
Shruti Kulkarni\IEEEauthorrefmark{1},
Lynn Tong\IEEEauthorrefmark{1},
Aditi Namboodiripad\IEEEauthorrefmark{1},
Chelyah Miller\IEEEauthorrefmark{1},\\
Helen Lin\IEEEauthorrefmark{1},
Peeyush Patel\IEEEauthorrefmark{1},
Bogdan Bistriceanu\IEEEauthorrefmark{2},
Diane Myung-kyung Woodbridge\IEEEauthorrefmark{1}
}
\IEEEauthorblockA{\footnotesize\IEEEauthorrefmark{1}MS in Data Science and Artificial Intelligence Program, University of San Francisco\\
skulkarni3@usfca.edu, latong@usfca.edu, anamboodiripad@usfca.edu,\\
cmiller9@usfca.edu, hlin65@usfca.edu, ppatel19@usfca.edu, dwoodbridge@usfca.edu}
\IEEEauthorblockA{\footnotesize\IEEEauthorrefmark{2}Accenture\\
bogdan.bistriceanu@accenture.com}
}

%% file: abstract.tex
\begin{abstract}
Wildfire seasons have become 84 days longer in the current days than in the 1970s, causing enormous threats to one’s financial status and short- and long-term health. During the fire, public agencies send out emergency messages to provide warnings and orders. Although 26 million people in the US have limited English proficiency, over 80\% of those messages are only delivered in English, which can cause disproportionate information distribution and awareness. In order to better serve marginalized communities during emergencies, the authors developed BEACON (Broadscale Evacuation Agentic Coordination, Outreach \& Navigation), a service that provides comprehensive and personalized evacuation guidance, including navigation routes, personalized checklists, and a chatbot in the language that a user uses. Our current system ingests data including fire perimeter information, evacuation order status, and shelter information from Watch Duty, a non-profit organization that provides live fire incident information via human monitoring, radio scanners, government data, satellite, etc.  When a user is within a certain proximity from the fire, the system utilizes real-time GPS locations and nearby weather data from the National Oceanic and Atmospheric Administration (NOAA) to predict fire danger levels. The assessment model refreshment and recalculation are dynamically scheduled based on fire progress and trends using XGBoost. If the location has a likelihood of fire danger, the system sends alerts with evacuation routes outputted from a polygon-avoidant routing pipeline to avoid any points and regions with fire danger. The application provides a context-aware multilingual agent that users can communicate with and is tightly connected/aligned to other features of the application. In addition, based on data that the user entered, the system dynamically generates and checks off personalized reminder items to provide an organized evacuation plan. The system’s user interface dynamically changes its language settings based on the language the user most recently used in either setting or chatbot conversation for all the application elements. The preliminary experiments showed the reliability and effectiveness of the developed features.
\end{abstract}

%% file: introduction.tex
\section{Introduction}
\label{sec:Introduction}
Climate change has extended the U.S. wildfire season by an average of 84 days compared to the 1970s, causing destructive outcomes. Recent fire incidents, including Camp (2018), Marshall (2021), Lahaina (2023), Palisades and  Eaton (2025), showed that timely evacuation has become harder with dry and windy climate conditions, causing a rapid spread of fire. In emergencies, it is critical to send and receive status updates and guidance through the local government to reduce damages and save more lives \cite{synolakis2024wildfire}. 
Unfortunately, over 80\% of wireless emergency alerts and state or county-level evacuation orders are issued exclusively in English, although 25.7 million US residents have limited English proficiency \cite{english_alert}\cite{acs2023}. This can be a huge bottleneck for marginalized populations and their families who cannot read and comprehend information sent in English, making them vulnerable.
In addition to the language barrier, the authors found further operational discrepancies in evacuation support systems. First, most commercial GPS services return evacuation routes that cross regions and points that were explicitly set to avoid. This could be a problem if those include active or potential fire perimeters and evacuation areas. This is a known limitation controlled by GPS application providers to minimize travel time without considering the exclusion requests. Additionally, most checklists and guides published by public agencies are static text and do not adapt to individual household conditions, including elderly, children, pets, livestock, mobility needs, medical needs, or vehicle availability, etc. \\
BEACON (Broadscale Evacuation Agentic Coordination, Outreach \& Navigation), an end-to-end agentic mobile application, includes a data ingestion pipeline, danger assessment based on dynamic evaluation based on danger levels, polygon- and point-based routing, a multilingual large language model (LLM) chatbot agent, and a personalized checklist to serve each family’s unique needs. The data pipeline ingests a user’s GPS data, real-time fire-perimeter data from Watch Duty, and weather data from the National Oceanic and Atmospheric Administration (NOAA). The system applies an Hourly Wildfire Potential (HWP) wind-adjustment layer that takes the distance-based danger score and escalates when the High-
Resolution Rapid Refresh (HRRR) fields indicate hot, dry, gusty conditions blowing toward the user \cite{james2025hourly}\cite{dowell2022high}\cite{HighReso24:online}.  The polygon and point routing takes regions with fire and evacuation warnings/orders, providing an option to include potential drop-by places to pick up families and friends or get necessary items during evacuation. The agent binds with the routing feature and checklist to ensure smooth operations across features throughout the user session, incorporating contextual information and previous conversation histories.

%% file: system_overview.tex
\section{System Overview}
\label{sec:system_overview}

\begin{figure}
    \centering
    \includegraphics[width=\linewidth]{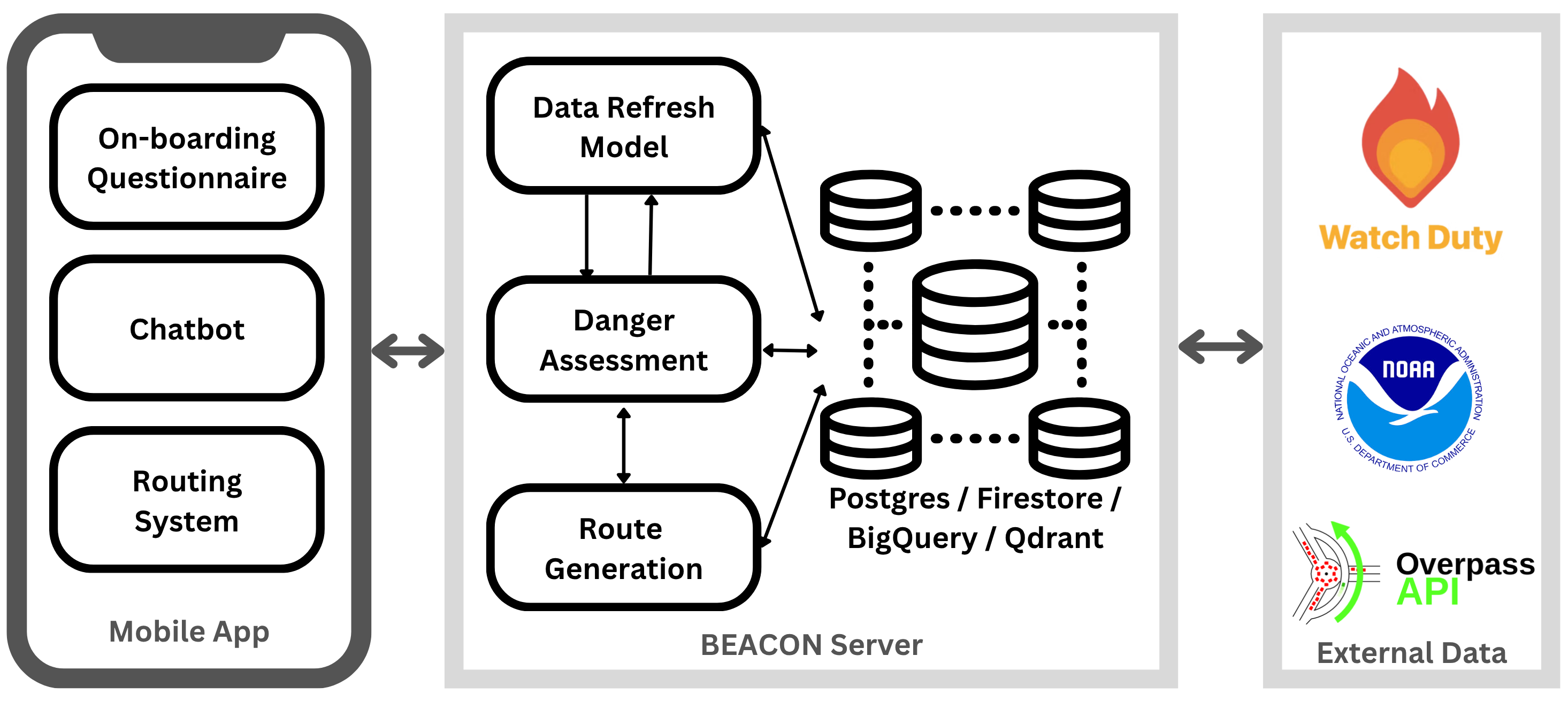}
    \caption{BEACON'S end-to-end data flow from a user's GPS ping through danger assessment, polygon-avoidant routing, and a multilingual agent that returns turn-by-turn evacuation steps and a personalized checklist.}
    \label{fig:system_architecture}
\end{figure}

The developed system includes a mobile application and backend server. The iOS mobile application was built using Swift, while the server is a containerized Python FastAPI service deployed on Google Cloud Run \cite{CloudCom78:online} (Figure~\ref{fig:system_architecture}).

%%%%%%%%%%%%%%%%%%%%%%%%%
\subsection{Data Ingestion}
\label{subsec:data_ingestion}
Using the mobile application's GPS location data, the system ingests fire perimeter and evacuation information from Watch Duty\cite{watchduty}, NOAA weather data for assessing fire danger, and points of interest from Overpass, including a chosen category by the user such as pharmacies, and widely used evacuation shelters including community centers, fairgrounds, etc. \cite{digrande2019pilot}. Those data are stored in the BigQuery database, rather than re-requesting via APIs.

If the user is within a threshold limit from fires or in evacuation areas, the system triggers a push notification to the user's mobile application and generates a route.

Additionally, during the sign-up step, users can provide a household profile using 12 binary questions, including elderly, children, pets, livestock, mobility needs, medical needs, or transportation availability, etc. Using the scraped guidelines, protocols, and existing checklist from various organizations, the system will generate a customized checklist for the user. Both family profile and customized checklist items are stored in a database.

%%%%%%%%%%%%%%%%%%%%%%%%%
\subsection{Danger assessment}
\label{subsec:danger_assessment}
If a user's location is not outside a threshold value for fire perimeters and evacuation zones, the system assesses danger levels. This danger level is re-assessed with a schedule determined by XGBoost classification.

The fire danger assessment uses an Hourly Wildfire Potential (HWP) score, computed on the HRRR grid from five surface fields including wind gust (\texttt{GUST}), 2-m temperature (\texttt{TMP}),
2-m dewpoint (\texttt{DPT}), soil moisture availability
(\texttt{MSTAV}), and snow water-equivalent (\texttt{WEASD}).

\begin{equation}
 HWP = 0.213 \cdot G^{1.50} \cdot \mathrm{VPD}^{0.73}
\cdot (1 - M)^{5.10} \cdot S,
\label{eq:hwp}
\end{equation}
where
\begin{itemize}
  \item $G$ : surface wind gust  ($\mathrm{m\,s^{-1}}$)
  \item $\mathrm{VPD}$ : 2-m vapor pressure deficit (hPa), computed from \texttt{TMP} and \texttt{DPT}
  \item $M \in [0,1]$ : soil moisture availability computed from $\texttt{MSTAV}/100$
  \item $S = \exp(-\texttt{WEASD}/10)$ : derived snow moisture value
\end{itemize}

The raw grid HWP is applied to a 9×9-sized kernel (approximately 27 km × 27 km) to reduce single-cell noise, and the user’s HWP is calculated from the nearest grid point using Euclidean distance between their longitude and latitude values.

\begin{equation}
            \widetilde{\text{HWP}}(i,j) = \frac{1}{81} \sum_{\Delta i=-4}^{4} \sum_{\Delta j=-4}^{4}\text{HWP}(i+\Delta i, j+\Delta j)
\end{equation}

The HWP score is used in the routing pipeline, where candidate routes whose length exceeds a configurable fraction $\phi_\mathrm{max}$ in cells with $HWP > \tau_{HWP}$ are discarded. 

As calculating HWP is time- and resource-consuming, we developed an adaptive refresh-cadence model to determine and schedule the next HWP calculation. The XGBoost algorithm uses previous HWP values, their momentum, and 24-hour mean values, and classifies the current location’s danger into safe, elevated, high, and extreme categories.

%%%%%%%%%%%%%%%%%%%%%%%%%
\subsection {Route Generation}
\label{subsec:route_generation}
While many existing GPS applications are a great choice for user adoption, the authors decided to embed a GPS feature within the application for two reasons. First, many existing platforms we evaluated offer does not provide features to render external fire-perimeter and evacuation zones as overlays. While the user sees the route, it is hard to know whether the route crosses a safe area during emergencies. Second, the existing platforms only consider points including departure, destination, and drop-by locations, but not points and polygons to avoid. To overcome the aforementioned limitations, the authors developed a GPS feature using OpenRouteService (ORS, \cite{ors}) which accepts a GeoJSON object as a hard constraint and returns a path guaranteed not to cross any supplied polygons and points.

The routing pipeline accepts a list of shelters from Overpass API and Watch Duty and, optionally, points of avoidance the user must avoid, such as reported accidents or road closures from the Overpass API.

The set of regions where the route should not cross is the union ($A$)
of active fire perimeter polygons, evacuation areas, and potential fire dangers from Section~\ref{subsec:data_ingestion} and Section~\ref{subsec:danger_assessment} and hazardous points.

The route is computed as a multi-stop path to a shelter where it has the shortest $W$ and does not pass $A$.

\begin{equation}
W \;=\; \bigl( o,\; w_1,\; w_2,\; \ldots,\; w_k,\; s \bigr),
\label{eq:waypoints}
\end{equation}
where $o$ is the user's GPS origin, $s$ is the candidate shelter,
and $w_1, \ldots, w_k$ are zero or more drop-by waypoints
selected by the user. The optimal routes and regions with dangers are rendered on the mobile application via Mapbox \cite{mapbox}.

However, when exhaustive searches do not find any destination with waypoints on the way, the system sends a message to the user and provides routes without waypoints. When the system cannot find a safe route, the agent triggers a push notification and sends an emergency instruction via chat.

\begin{figure}[t]
\centering
\begin{subfigure}[b]{0.49\linewidth}
    \centering
    \includegraphics[height=8.5cm,keepaspectratio]{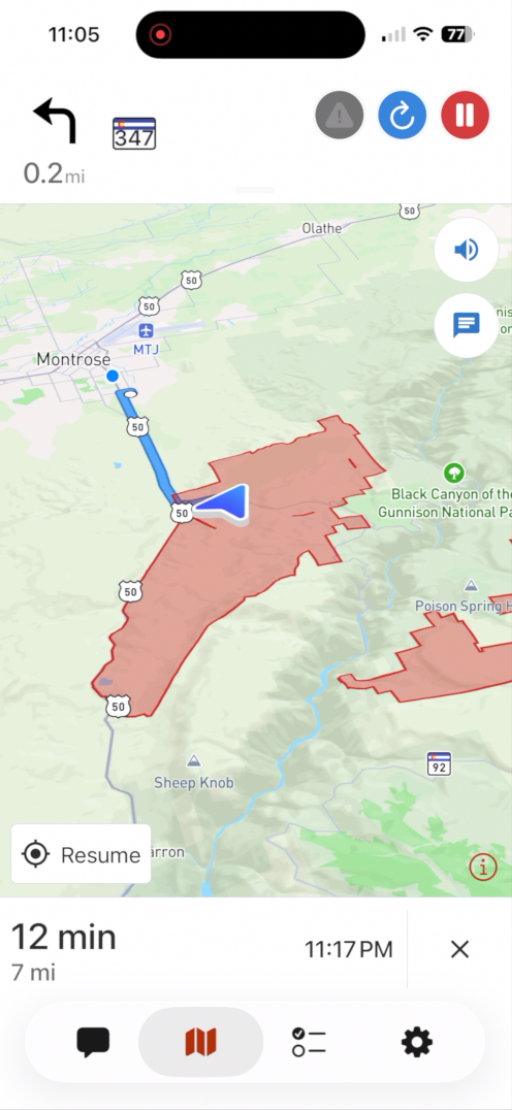}
    \caption{Route with fire}
    \label{fig:route_overlay}
\end{subfigure}
\hfill
\begin{subfigure}[b]{0.49\linewidth}
    \centering
    \includegraphics[height=8.5cm,keepaspectratio]{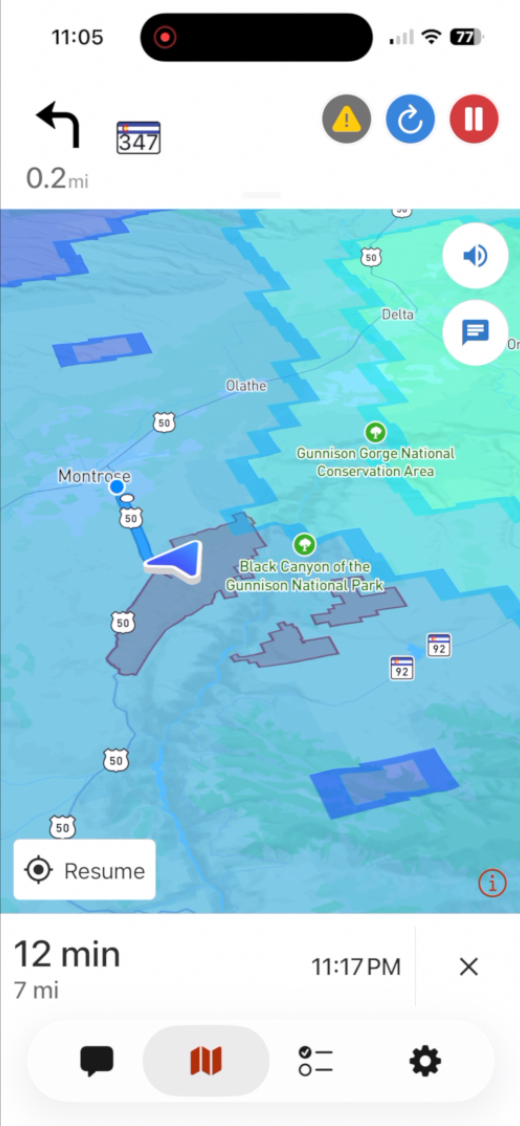}
    \caption{Route with fires and HWP}
    \label{fig:route_overlay_pwd}
\end{subfigure}
\caption{Rendered fire and HWP regions and safe routes}
\label{fig:routes}
\end{figure}

%%%%%%%%%%%%%%%%%%%%%%%%%
\subsection{The Multilingual Agent}
BEACON’s conversational feature is an LLM-driven agent, not a simple chatbot. The agent creates a six-block prompt for each session and manages three features, namely routes, checklist, and user preferences. Based on the keywords used in the chat or the classified intended uses, the agent chooses the feature and properly updates/share the status. For instance, if a user says ``I packed my prescribed medicine``, the corresponding item will be checked on the checklist.  

The prompt is generated at the beginning of each session, including the following six blocks.
\begin{enumerate}
  \item \textbf{Identity}: Statements of its role (``BEACON, a wildfire evacuation assistant'') and what it must and must not do (i.e, no clinical advice, no unverified information).
  \item \textbf{Location}: Raw coordinates of the user.
  \item \textbf{Active perimeters and urgency}: the active
        evacuation zones with the urgency levels.
  \item \textbf{Route status}: Either ``a safe route has been loaded
        and is visible on the map'' or ``no safe route could be
        computed. direct the user to call 911''.
  \item \textbf{Personalized checklist context}: Twelve household questions/answers and the list of checklist items and status.
  \item \textbf{Language rules}: Most recently used language by the user to ensure a response in the same language.
\end{enumerate}

When the agent detects changes in the languages that the user used in chat or manually set in the user preferences, this new status is persisted in the database and application. This setting not only affects the language used in the chat, but is also applied to the application UI, including buttons/preference settings, route instructions, and checklist.

In addition, BEACON maintains Qdrant as an embedding storage for cross-session memory \cite{QdrantVe57:online}. This helps the agent to add recent conversation summaries and the user profile and injects them as ``what I know about the user'' for a smoother conversation.

\begin{figure}[t]
\centering
\begin{subfigure}[b]{0.49\linewidth}
    \centering
    \includegraphics[height=8.5cm,keepaspectratio]{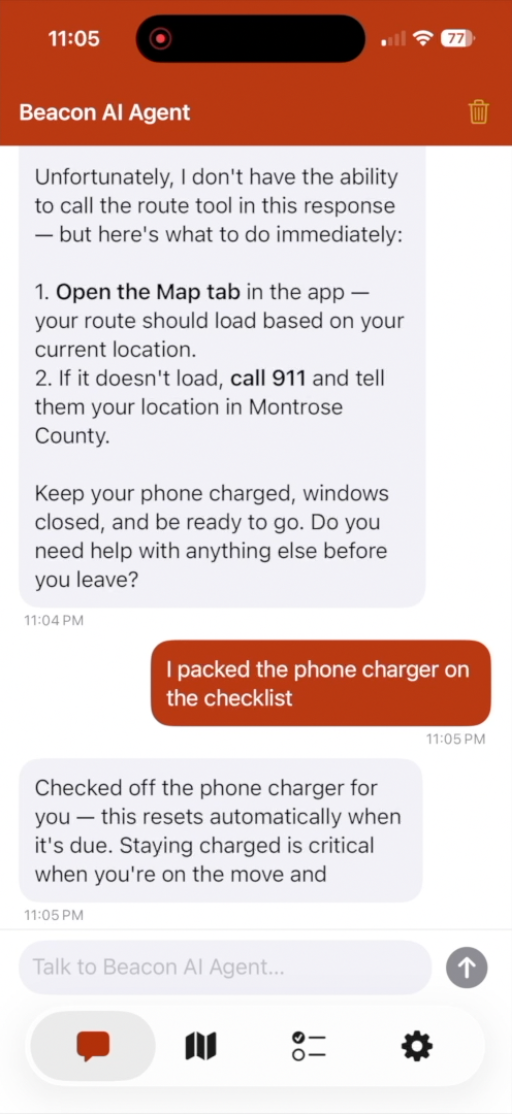}
    \caption{Chat (English)}
    \label{fig:chat_eng}
\end{subfigure}
\hfill
\begin{subfigure}[b]{0.49\linewidth}
    \centering
    \includegraphics[height=8.5cm,keepaspectratio]{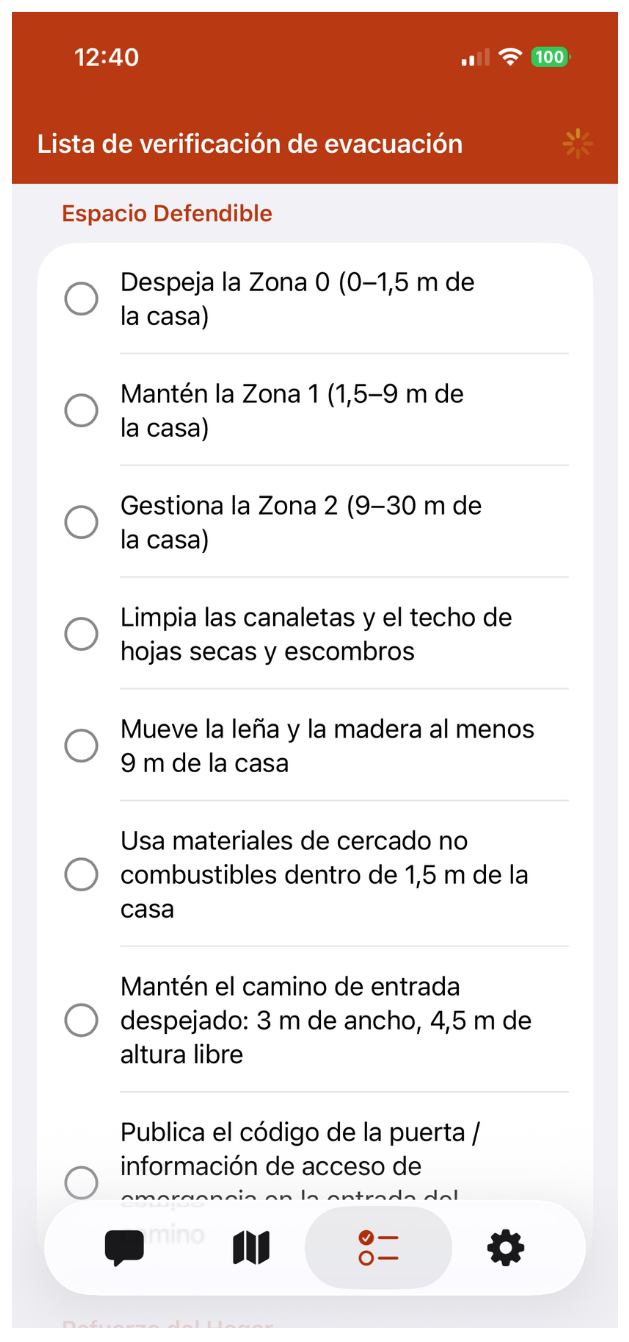}
    \caption{Checklist (Spanish)}
    \label{fig:chat_span}
\end{subfigure}
\caption{BEACON's Multilingual agent}
\label{fig:language}
\end{figure}

%% file: results.tex
\section{Results}
\label{sec:results}

The results below are qualitative observations of the end-to-end system behaviors, focused on the four properties we set out to deliver, including automatic language detection without configuration, polygon-avoidant routing under realistic fire perimeters, context-aware conversation, and graceful degradation when one or more inputs are unavailable.

For this experiment, we concentrated on Colorado from July to September 2025, where there were nine fires, each burning more than 1,000 acres, using data provided by Women in Data Science Worldwide \cite{WiDSUniv69:online}.

\subsection{Automatic language detection without configuration}
Across evaluation sessions,  the agent reliably identified the user's language from the first message and replied in the same language without manual selection. Furthermore, the entire user interface on the mobile application was displayed in the user's language.\\
We also verified that the agent honors mid-session switches, where the agent followed each switch from the most-recent message (Figure~\ref{fig:language}).

\subsection{Polygon-avoidant routing under realistic perimeters}
With Watch Duty polygons of an active perimeter near Montrose County, Colorado, the polygon-avoidant routing pipeline consistently returned paths that avoided every active fire polygon
(Figure~\ref{fig:routes}). For the same input geometry, the
baseline commercial routing APIs in our preliminary testing returned paths that crossed regions explicitly requested for being excluded.

\subsection{Context awareness from the first turn}
Because the system prompt assembled at session start contains the
danger category, the loaded safe route, the household profile, and the previous conversation summary, the agent responds with full context on the very first user turn. In evaluation sessions, a user's opening message ``Should I leave?'' at the beginning of a session produced a tailored response that referenced the user's distance to the nearest perimeter, their pre-loaded route, and household-specific items without the user having to repeat.

\subsection{Graceful degradation under operational failures}
A core design principle of the system is that no single
failure should cause the system failure for a user who may be in immediate danger. Each external data and service should either continue with a degraded but still provide a response. We summarize the patterns tested and validated.

\begin{itemize}
  \item \textbf{No safe route.} When the routing pipeline
        exhausts all shelter candidates, the agent's response switches to a ``call 911 now'' message, and a push notification is dispatched.

  \item \textbf{Drop-by constraint infeasible.} If a
        user-requested waypoint are not within the route's proximity to ant shelter candidates, the pipeline retries without the drop-by and share the route with the user informing that the waypoint was not available.

  \item \textbf{Routing language unsupported.} If the user's
        preferred language is not in the 23 languages
        natively supported by ORS, the route is requested in
        English and the turn-by-turn instructions are translated
        via Claude, so the user still communicate in their own
        language.

  \item \textbf{Semantic classifier failure.} If the Layer-2
        GPT call fails or times out, the per-turn intent
        defaults to \texttt{Intent.NONE} and the agent responds
        from the system prompt alone, without tool exposure.
        Latency on these turns is bounded by the keyword pass
        (Layer~1) plus the failed-API timeout.

  \item \textbf{Cross-session memory unreachable.} If the Qdrant
        retrieval errors or returns no results, the
        \emph{What I Know About This User} block is omitted from
        the system prompt and operates from the per-session
        context only.

  \item \textbf{Missing household onboarding.} When the user has
        no household information, the personalized
        checklist context is omitted from the system prompt and
        encourage the user to complete the twelve household questions in the iOS Checklist tab.
\end{itemize}

In every observed degraded session, the agent continued to provide
some useful response, and the escalation paths to emergency services were preserved.

%% file: conclusion.tex
\section{conclusion}
\label{sec:conclusion}
In this preliminary research and development, we developed a multilingual agent to provide safe routes and advice to communities that have limited English during a fire. The developed system ingests data from the user and external sources, including Watch Duty and NOAA, and assesses danger levels for the user. The reassessment will be scheduled dynamically based on the recent HWP values and trends.
The qualitative experiment demonstrated the reliability and effectiveness of the implemented features.

In future research, we will incorporate other APIs and databases for fires and live shelter availabilities. We are also planning to develop an Android application to reach more communities. Additionally, a speech-to-text feature could enhance user experience while driving.